\documentclass[iop]{emulateapj}

\usepackage{rotating}
\usepackage{threeparttable}

\shorttitle{Correlations between SNe Ia Rates and Host Galaxy Properties}
\shortauthors{Gao and Pritchet}

\usepackage{amsmath}
\usepackage{amsthm}
\usepackage{amssymb}
\usepackage{dcolumn}
\usepackage{longtable}
\usepackage{multirow}

\begin{document}


\title{Correlations between SDSS Type Ia Supernova Rates and Host Galaxy Properties}


\author{Yan Gao\altaffilmark{1} \and Chris Pritchet}
\affil{Department of Physics \& Astronomy, University of Victoria}
\email{gaoy@uvic.ca}
\altaffiltext{1}{School of Astronomy and Space Sciences, Nanjing University}



\begin{abstract}
Studying the correlation of type Ia supernova rates (SNR) with host
galaxy properties is an important step in understanding the exact
nature of type Ia supernovae. We use SNe Ia from the SDSS-II sample,
spectroscopically determined masses and star formation rates, and a
new maximum likelihood method, to fit the Scannapieco and Bildsten
rate model $SNR = A{\times}M + B{\times}SFR$, where $M$ is galaxy mass
and $SFR$ is star formation rate. We find $A =
3.5^{+0.9}_{-0.7}\times10^{-14}(\text{SNe/yr})(M_{\odot})^{-1}$ and $B
=
1.3^{+0.4}_{-0.3}\times10^{-3}(\text{SNe/yr})(M_{\odot}$yr$^{-1})^{-1}$,
assuming overall efficiency of 0.5. This is in reasonable agreement with other determinations. However we find
strong evidence that this model is a poor fit to other projections of
the data: it fails to correctly predict the distribution of supernovae
with host mass or SFR. An additional model parameter is required; most
likely this parameter is related to host galaxy mass. Some
implications of this result are discussed.
\end{abstract}





\section{Introduction}

Supernovae of type Ia (SNe~Ia) played a pivotal role in the discovery
of the accelerating Universe (Riess et al.~1998, Perlmutter et
al.~1999), and are among the most reliable standard candles on
cosmological distance scales (Sullivan et al. 2011, Barone-Nugent et
al. 2012). However, a complete understanding of SN Ia progenitor
systems has yet to be achieved, even though credible models for such
progenitors exist (e.g. Hillebrandt \& Niemeyer 2000, Nomoto 1982,
Hachisu et al.~1996). Many methods have been developed to discriminate
between proposed models, including the theoretical modelling of
progenitor systems (e.g., Han \& Podsiadlowski 2006, Wang et
al.~2010), and the measurement of delay time distributions (e.g.,
Mennekens et al. 2010, Totani et al. 2009, Sand et al.~2012; Maoz et al.~2012; see also
Greggio 2005, 2010).

One approach used in constraining the nature of SNe~Ia progenitors is
studying the correlation between SN~Ia rates (SNR) and the properties
of the galaxies in which they form (e.g., Maoz \& Mannucci 2011). The
environmental properties studied so far include redshift (e.g., Dilday
2010a and 2010b), host galaxy age (e.g., Gupta et al.~2011),
environment galaxy number density (e.g., Cooper et al.~2009), ejecta
velocity (e.g., Foley 2012), and, most relevant to this work, mass and
star formation rate of the host (e.g., Sullivan et al.~2006).

In this paper, we investigate the correlation between SN~Ia rates and
the mass and star formation rate of the host galaxies in which they
formed. One popular model for this correlation is the ``$A+B$" model
(Scannapieco \& Bildsten 2005), which states that the supernova rate
is proportional to a linear combination of galaxy mass and star
formation rate: $SNR = A{\times}M + B{\times}SFR$, where $SNR$ is the
SN~Ia rate, $M$ is galaxy mass in stars, and $SFR$ is star formation
rate. This relation leads trivially to a more physical relation $SNR/M
= A + B{\times}(SFR/M)$, where $SNR/M$ is the rate per unit mass, and
$SFR/M$ is specific star formation rate (which is a function of star
formation history, independent of mass).

Substantial effort has been invested in determining the values of $A$
and $B$ since the work of Scannapieco and Bildsten (2005). See
Fig. \ref{litcomp} for a comparison of some of these results. Neill et
al. (2006) and Dilday et al. (2008) used volumetric SN~Ia rates to
infer these parameters. Of particular interest is the work of Sullivan
et al.~(2006), who obtained stellar masses and SFRs of each individual
host (and field) galaxy by fitting PEGASE2 spectral templates to
multiband photometry (Fioc \& Rocca-Volmerange 1997, Le Borgne and
Rocca-Volmerange 2002). They concluded that
$A=5.3\pm1.1\times10^{-14}(\text{SNe/yr})(M_{\odot})^{-1}$,
$B=3.9\pm0.7\times10^{-4}(\text{SNe/yr})(M_{\odot}$yr$^{-1})^{-1}$,
values that we will return to.

On the other hand, Smith et al.~(2012) used photometric masses and
SFR's to demonstrate that the $A+B$ model was not a good match to
SDSS-II supernova data (Frieman et al.~2008). They proposed an
alternate model of the form $SNR = A{\times}M^{x}+B{\times}SFR^{y}$,
with $x\simeq 0.7$ and $y\simeq 1$. This is, on the surface, broadly
consistent with the mass dependence found by Li et al.~(2011) from the
Lick Observatory Supernova Survey (LOSS) (though see \S 6).

The primary objective of this work is to obtain fits to SNR models
using \emph{spectroscopically}-determined masses and SFRs from the
SDSS DR7 MPA/JHU value-added catalog (Kauffmann et al.~2003,
Brinchmann et al.~2004, Tremonti et al.~2004), making these values
somewhat independent of, and possibly more reliable than, those based
on multiband photometric data. This is of significance since it has
been pointed out (e.g., F\"orster et al.~2006) that the uncertainty in
measurements of star formation histories is an important limiting
factor in the determination of the values of $A$ and $B$. In addition,
we use a new fitting algorithm for $A$ and $B$ based on maximum
likelihood, which is more reliable than previous fitting methods.

In \S 2, we present the data samples we use in our studies, which
include the sample of SNe~Ia, the sample of photometric galaxies, and
the subset of spectroscopic galaxies (itself a subset of the
photometric sample). Also presented in \S 2 is the essential
preprocessing of the data, most notably the matching of the SNe~Ia to
their host galaxies in the photometric sample, and a brief
introduction to the MPA/JHU masses and SFRs, in addition to a
comparison with other masses and SFRs. Our results for $A$ and $B$ are
given in \S 3, and a detailed examination of the fit of the model to
the data is given in \S 4. Finally we discuss alternate models in \S
5, and the implications of these alternate models in \S 6.

\section{Data}

The Sloan Digital Sky Survey (SDSS, Abazajian et al.~2009) Data
Release 7 catalog\footnote{http://www.sdss.org/dr7/} (DR7) contains
357 million unique objects, of which nearly 930,000 are galaxies for
which spectra are available. From these spectra, masses and star
formation rates have been obtained (Kauffmann et al.~2003, Brinchmann
et al.~2004, Tremonti et al.~2004, hereafter MPA/JHU); see \S 2.2 for
details. Supernova identifications were taken from the SDSS-II
Supernova
Survey\footnote{http://sdssdp62.fnal.gov/sdsssn/snlist\_confirmed\_updated.php},
which is a 3-year (2005-2007) survey conducted within Stripe 82 of the
SDSS. Stripe 82 covers an area of nearly 300 square degrees in a belt
along the celestial equator, and contains $\sim$20,000 of the
spectroscopic galaxies mentioned above. For each year the survey was
conducted, supernova imaging was conducted for 3 months, resulting in
a total observation period of 9 months.

\subsection{SDSS II Type Ia Supernovae}

From the complete sample of 660 SDSS II supernovae (Frieman et
al.~2008), we extract 520 spectroscopically confirmed SNe~Ia, 503 of
which were observed during the 3 observation seasons in 2005, 2006 and
2007. (For the rest of this paper, ``SNe" refers to SNe~Ia unless
stated otherwise.) Our sample is larger than that used by Smith et
al.~(2012), since we include SNe within $z=0.05$. We omit 17 SNe
observed in 2004, since the observation windows and completeness of
these SNe would be hard to gauge.

Dilday et al.~(2008, 2010) have shown that the identification
efficiency $\epsilon_{z}$ in the SDSS-II Supernova Survey is $\gtrsim
0.7$ out to a redshift of 0.15, and $>$0.5 out to z=0.25 (e.g. Fig. 8 of Dilday et al.~2010).
In fact the median z of SN in our (spectroscopic host) sample is much lower ($\langle z \rangle = 0.11$) than for the SDSS SN Ia sample as a whole ($\langle z \rangle = 0.2$) because of the severe magnitude limit $r=17.77$ for the spectroscopic sample of hosts. Thus incompleteness $\epsilon_z=0.7$ appears to be a reasonable assumption. We also assume
an observation window $T= 9$ months in
length, and detection efficiency $\epsilon_{t}=1$. The size of
the SN sample is therefore assumed to be the intrinsic number occuring in the
sky within a timeframe of $\epsilon_{t}\epsilon_{z}T=0.5$ year. This
assumption is discussed further in \S 6.1, but we note here that
different values of $\epsilon_{t}\epsilon_{z}T$ result in a simple
scaling of supernova rates, and hence of $A$ and $B$ values.

\subsection{SDSS Spectroscopic Masses and Star Formation Rates}

The SDSS DR7 MPA/JHU value-added catalog contains derived masses and
SFR values for the spectroscopic galaxies, including 19987 galaxies in
Stripe 82 (our spectroscopic sample). The mass estimates follow a
methodology similar to that of Kauffmann et al.~(2003), namely a grid
search over a library of star formation histories (Bruzual \& Charlot
2003) for the most probable mass-to-light ratio. (This is slightly
different from the approach of Kauffmann et al.~(2003), which matches
$D_{n}(4000)$ and $H\delta_{A}$, but it has been shown that the mass
estimates of the two methods\footnote{http://www.mpa-garching.mpg.de
/SDSS/DR7/mass\_comp.html} do not differ by more than $\sim 0.1$dex.)
The methods used for obtaining SFR measurements were similar to those
proposed by Brinchmann et al.~(2004). Galaxies were divided into three
classes: (1) ``SF", (2) ``low S/N SF", and (3) ``AGN, Composite and
Unclassifiable". The ``SF" class was processed by fitting 5 emission
lines ($H_{\alpha}$, $H_{\beta}$, O III, N II and S II) to those of
simulated galaxies in a spectral library obtained using the code by
Charlot \& Longhetti (2001). For the ``low S/N SF" class, Brinchmann
et al. used a simple conversion factor between attenuation-corrected
$H_{\alpha}$ luminosity and SFR. The ``AGN, Composite and
Unclassifiable" class was processed by means of the empirical relation
between SFR and D4000.

One problem with this methodology is that passive galaxies are always
assigned a (very small) SFR. This can be seen in Fig. \ref{fig2_1},
where MPA/JHU spectroscopic galaxies are plotted in the mass-SFR
plane. It is demonstrated in \S4 that this does not significantly
affect our results.

To test the reliability of the MPA/JHU masses and SFR's, we compare
their results with those of the VESPA catalogue (Tojeiro et al.~2009),
which estimates the star formation history of a number of galaxies in
SDSS Stripe 82 using spectroscopic methods. VESPA masses are obtained
by summing the star formation history; VESPA SFR's are obtained from a
SFH averaged over the redshift range 0 to 0.11. A comparison of these
quantities with their MPA/JHU counterparts is made in
Fig. \ref{fig2_2}. From this figure, it can be seen that the mass and
SFR estimates from the two catalogues are in reasonable agreement with
each other, with offsets of 0.25 and 0.24 for log M and log SFR, and
rms scatter of 0.18 dex and 0.48 dex, respectively.

At this point it is relevant to ask how much better
spectroscopically-determined masses and star formation rates are than
their photometric counterparts (as used by, for example, Sullivan et
al.~2006, and Smith et al.~2012). The advantage of using spectroscopic
data is due to several effects.  1. Redshift estimates (required for
both the SN and field samples -- e.g., \S 3) are much better for
spectroscopic data.  Using the same sample as ours, Smith et
al. (2012) showed that photometric redshifts result in negligible offsets
(0.03 dex) but a scatter of $>0.2$ dex in log M and log SFR relative
to using spectroscopic redshifts.
2. Photometric methods for obtaining M and SFR from SDSS data rely on
fitting evolving stellar population models to 5 broad photometric
bands. The simplicity of these models, coupled with the wide range of
degeneracies between model parameters, will result in larger random
errors, particularly for SFR. By comparison spectroscopic methods
allow the use of far more data, including emission lines, absorption
lines, and spectral breaks that are better tuned to the determination of M
and SFR.
3. From a comparison of two different catalogs above, we have already
seen that there are significant systematics in the determination of
spectroscopic M and SFR.  For all the reasons mentioned in point 2,
there are likely to be even larger systematic errors in photometric M
and SFR measurements. These systematics are of the greatest concern; none of
our conclusions are affected by random errors in M and SFR.


\subsection{Host Matching}

We adopt the host matching algorithm of Sullivan et al.~(2006). We use
SDSS DR7 $r$ band data to calculate a dimensionless parameter $R_{25}$
for every potential host - SN pair. $R_{25}$ is the angular separation
of a SN from its prospective host, measured in units of the size of
the 25 magnitudes/arcsec$^{2}$ isophote, and allowing for the shape of
this isophote (see Fig. \ref{R25}). The host candidate is then
identified to be the galaxy with the lowest $R_{25}$. SNe~Ia with
redshifts falling more than $3\sigma$ from the photometric redshift of
the host galaxy are then discarded. Using such a selection procedure, for hosts with spectroscopy, host
z and SN z are always in agreement to a precision of 0.001.

How large can $R_{25}$ be without introducing significant
contamination? To answer this question, we conducted Monte Carlo
simulations with randomly distributed artificial SN positions, keeping  in mind that a  majority of contamination is due to SNe for which the host is not seen. We find
that the contamination is $\sim$ 8\% when only SN-host candidate pairs
with $R_{25}<3.8$ are identified as genuine matches. Contamination
could be larger if we were to take clustering effects into account,
but such effects are hard to quantify. Applying this criterion, we
find 351 matches for the SNe within the SDSS DR7 database. Of these,
53 are galaxies in the spectroscopic sample, with MPA/JHU masses and
SFRs. It may be argued that since $\sim$ 170 SNe were unmatched,
$\sim$ $8\%\times170\sim14$ SNe of the 351 could be random matches. We
consider this number to be relatively small in comparison to the 351
matches. Also, this contamination estimate corresponds to the expected
contamination from using $R_{25}$ alone as matching criteria. The
additional redshift constraints would help further lower the number of
random matches. The question of contamination is re-examined in \S 6.2.

To test the host-matching procedure above, we use 2 separate
methods. The first is to visually examine the images of the SN - host
pairs, relative to their $R_{25}=1$ isophotes. The second is to plot
the number of SN matches versus the $R_{25}$ criterion adopted
(Fig. \ref{fig2_6}). We can see that there is a flat region beyond
$R_{25}=3.8$, which we assume to be the regime where random matches
between SNe and non-host galaxies become dominant.

We also compare the discrepancies between our host-matching method and
the ``closest angular distance" matching method used by many other
authors (e.g. McGee and Balogh 2010). We find that the two matching
methods give different results for $\sim$10\% of the SNe.

\section{Fitting $A$ and $B$}

We adopt a fitting algorithm based on maximum
likelihood. Using Poisson probabilities, the likelihoods of a galaxy
hosting/not hosting a SN~Ia are \begin{equation} P(S_{i}|SN)=S_{i}
e^{-S_{i}}, \text{\ and\ } P(S_{i}|\overline{SN})=e^{-S_{i}},\\
\end{equation} \noindent where $S_{i}$ is the expected SN~Ia rate for
each galaxy within our observing window, calculated for every galaxy
individually by means of the mass and SFR data. (None of the hosts in our sample hosted more
than one SN.) Multiplying these
likelihoods for every galaxy, we obtain a log likelihood of obtaining
our dataset of
\begin{equation}
{\rm ln} L=-\displaystyle\sum_{n=1}^{19987}S_{n}+\displaystyle\sum_{m=1}^{53}({\rm
ln}S_{m})+\text{constant}.
\end{equation}
\noindent where 53 is the
number of galaxies hosting SNe, and 19987 is the total number of
galaxies in our Stripe 82 spectroscopic sample.

We perform a grid search to find the maximum likelihood $A$ and $B$
values, which are
$A=3.5^{+0.9}_{-0.7}\times10^{-14}(\text{SNe/yr})(M_{\odot})^{-1}$ and
$B=1.3^{+0.4}_{-0.3}\times10^{-3}(\text{SNe/yr})(M_{\odot}$yr$^{-1})^{-1}$. Fig. \ref{fig3_1}
plots the probability contours and error bars for our grid search,
while Fig. \ref{fig3_2} compares these results with the observed SNR
at different specific SFR (where specific SFR $sSFR=SFR/M$)

While exceptionally simple, the maximum likelihood analytical method
above does have the drawback that it neglects statistical uncertainties in the
determination of M and SFR. The method could be modified to include
the effects of these uncertainties, however, given our conclusion
(\S4) that the A+B model is not valid, we have chosen instead to
simply simulate the effects of these uncertainties. We have conducted a set
of Monte Carlo simulations by seeding random, artificial SNe into the
spectroscopic sample, and have included the effects of uncertainties in M and
SFR. We find that A and B are biased by +0.05 and -0.05 when the
effects of uncertainties in M and SFR are included. (This bias is in the
sense ``derived value minus true value''.) The mass term and the SFR
term contribute 27 and 26 SNe Ia respectively, implying that there are
as many prompt SNe Ia as delayed SNe Ia according to the model.

Our final results
($A=3.5^{+0.9}_{-0.7}\times10^{-14}(\text{SNe/yr})(M_{\odot})^{-1}$
and
$B=1.3^{+0.4}_{-0.3}\times10^{-3}(\text{SNe/yr})(M_{\odot}$yr$^{-1})^{-1}$)
are consistent within error bars with most of the literature. For a
summary/comparison of these results, see Fig. \ref{litcomp}.

\section{Inconsistencies in the $A+B$ Model}

The $A$ and $B$ values above are optimal in a maximum likelihood
sense. However, there is no guarantee that the $A+B$ model provides a
good fit to our data; more tests are needed.

As a test of the $A+B$ model's consistency, we rank the spectroscopic
galaxy sample of 19987 galaxies by mass, calculate the cumulative
percentage distribution of SNR for both the $A+B$ model and the
observations, and apply a Kolmogorov-Smirnov (K-S) test to the two
distributions (Fig. \ref{fig4_1}). The $A+B$ model with our derived
values for $A$ and $B$ is rejected at the 99\% confidence level. This
result was also found by Smith et al.~(2012).

To investigate the possibility that other $A$ and $B$ values could
have passed the test, we plot (Fig. \ref{fig4_1}) the cumulative
distribution functions of our galaxies for both an $A$-only model
(SNR=$A{\times}$M) and a $B$-only model (SNR=$B{\times}$SFR), keeping
in mind that any set of values of $A$ and $B$ must fall between these
two distributions. (Note that for the $A$-only and $B$-only models,
the values taken for $A$ and $B$ do not affect the result of the K-S
test.) The $A$-only model is rejected at an even higher degree, while
the $B$-only model passes the test ($\sim$ 50\% rejection). This shows
that our data do not support a larger value of $A$/$B$.

Passing the mass-ranked K-S test is a necessary but insufficient
condition for a SNR model to be plausible. Almost all models pass the
\emph{sSFR-ranked} K-S tests, including the $A+B$ model
(Fig. \ref{fig4_7}), but the same cannot be said for the SFR-ranked
K-S tests, for which both the generic $A+B$ model, the $B$-only model,
and all models in between are rejected (Fig. \ref{fig4_8}). This means
that a smaller value of $A$/$B$ will not pass the test. Since we
already know that a larger $A$/$B$ value will not work (see above), we
conclude from these tests that \emph{no $A+B$ model can match our
observed data}. The results of these KS tests are shown in Table
\ref{ABalpha}.

The KS test is usually valid only when the measurement uncertainties in the observed sample are small. Therefore the (possibly significant) uncertainties in the mass and SFR measurements brings the validity of our above methods into question. In order to check the validity of our K-S test under such conditions, we use a Monte
Carlo approach. We generate many artificial samples of supernovae
using the best fit A and B values with the sample of 19987 observed M
and SFR values, add the effects of uncertainties in M and SFR, and generate
cumulative distributions of log M and log SFR. The results are
completely consistent with the K-S tests: the A+B model is rejected at
$>$99\% probability for log M and $>$97\% probability for log SFR.

The maximum likelihood method for fitting A and B returns a
probability which can be used as a measure of goodness of fit. This
probability can be interpreted using Monte Carlo simulations. Allowing
for the fact that this probability depends on the number of
supernovae generated (which varies from run to run), we find that more
than 99.9\% of all simulations possess a peak probability greater than
that of the best fit to the observations. Thus we conclude,
independently of the K-S tests, that the A+B model is not a good match
to the data.

To present an alternate, and somewhat more illustrative, view of this
discrepancy, we separate the galaxies by mass into two equally-sized
groups divided at log M=10.7, and plot specific SNR vs. specific SFR
(as in Fig. 6 of Sullivan et al.~2006). The resulting plot
(Fig. \ref{fig4_2}) shows that \emph{low-mass galaxies have a
systematically higher specific SNR than their high-mass counterparts,
by a factor of 3--5.} Applying different mass cuts consistently
results in the same trend.

A plot of the differential distribution functions of SNR predicted by
the $A+B$ model vs the actual observed SNR (Fig. \ref{fig4_3}) further
illustrates the issue, where the curves have been scaled to show the
relative absolute SNR obtained by each model. From Fig. \ref{fig4_3},
it can be seen that both the predictions of our best-fit $A+B$ model
(green line) and that of Sullivan et al.~(2006) (blue line)
underpredict the rates of supernovae hosted by low mass, high specific
star formation rate galaxies, while overpredicting the rates in high
mass galaxies. The red line is a (somewhat unphysical) modified $A+B$
model with a constant background supernova rate added (\S 4.1). The
two (unmodified) $A+B$ models are in reasonable agreement with each
other.

To show how large an effect our uncertainties in the determination of the $A$
and $B$ parameters may have on Fig. \ref{fig4_3}, we plot
Fig. \ref{ABerr}, which shows the differential distributions for the
1$\sigma$ upper and lower limits for $A$ and $B$. It can be seen that
the uncertainties cannot account for the discrepancy at the low mass end.

As mentioned in \S 2, the MPA/JHU catalog systematically overestimates
the SFRs of passive (non-star-forming) galaxies, assigning them a
small but significant SFR. We redo the $A+B$ fits by setting the SFRs
of all galaxies with an sSFR smaller than $10^{-11.5}$
$(M_{\odot}/\mathrm{yr})/M_{\odot}$ to zero; our results are
unaffected.

\section{Modifications to the $A+B$ Model}

We have demonstrated that the $A+B$ model does not reproduce our
data. To account for the discrepancy, most notably the observed SNR
excess in low mass galaxies, we now try modifications to the $A+B$
model. Table \ref{ABalpha} summarizes the results for some of the many
models tried. Due to our small sample size, and also to intrinsic
uncertainties in the MPA/JHU measurements of masses and star formation rates,
there exist strong correlations between model parameters when more
than 2 parameters are used.

As previously found by Sullivan et al.~(2006), $A\times M$- and $B\times SFR$-only
models do not match the observed SNR distributions. The
$A{\times}M+B{\times}SFR+C$ and
$(A{\times}M+B{\times}SFR)(1+C{\times}M^{-1})$ models are particularly
interesting, since they both pass the KS tests, and might have a
physical explanation (\S 6).

We also investigate the model proposed by Smith et al.~(2012): $SNR =
A{\times}M^{x}+B{\times}SFR^{y}$.  We attempt to obtain best-fit
values for the parameters of our own using maximum likelihood, but
find that our sample is too small to overcome the correlations between
4 parameters. Using Smith et al.'s values of $x=0.72$, and $y=1.01$,
we refit $A$ and $B$ using maximum likelihood; the result is
$A=5.3^{+1.3}_{-0.9}\times10^{-11}, B=1.0^{+0.4}_{-0.3}\times10^{-3}$
(appropriate units in the $M_{\odot}$ - yr system). Using either these
values, {\it or} the $A$ and $B$ values given by Smith et
al. ($A=0.41\times10^{-10}, B=0.65^\times10^{-3}$), we find that the
model passes the mass-ranked test, but fails the SFR-ranked test at
99\% confidence. A very small value of $x=0.06\pm0.10$ fits the data
better, and passes both KS tests - this is probably a manifestation of
the C term mentioned above.

\section{Discussion}\

In this paper, we have obtained $A$ and $B$ values from spectroscopic
data. However, further analysis reveals that there is an intrinsic
inconsistency between the data and the $A+B$ model: mass-ranked KS
tests reject all $A$ and $B$ values with a higher $A$/$B$ ratio than
our best-fit values, while SFR-ranked KS tests reject those with a
smaller $A$/$B$ ratio; furthermore, from Figs. \ref{fig4_1},
\ref{fig4_2} and \ref{fig4_3}, we can see that there exists a SNR
excess in low mass galaxies. \S 6.1 discusses some possible issues
with the data and its interpretation; \S 6.2 interprets our results.

\subsection{Data Issues}

We have not applied any completeness corrections to the
supernovae. The completeness of the sample has been discussed in \S2;
incompleteness will affect the numerical values of $A$ and $B$ by less
than 0.1 dex, but will not affect $A$/$B$, nor any of the conclusions
regarding the $A+B$ model (unless there is strong redshift evolution
in the SNR -- something that is not observed -- e.g., Perrett et
al. 2012). To check this, we redo the work with a redshift cut of 0.20
instead of 0.25, and find no significant differences.

Is it possible that the detection efficiency is different in galaxies
with high star formation, and concomitant higher extinction? We have
already noted that the detection efficiency of our sample is higher
than for SDSS SNe as a whole (\S 2.1). In fact high specific SFR
implies lower M and luminosity (Brinchmann et al.~2004) and as a result the star
forming galaxies (log sSFR $> -10.5$) in our sample have an even lower
mean z (about 0.07); this means that incompleteness is even less of a
problem. A simulation of an extreme efficiency variation with sSFR
shows that, while $B$ will be biased, the main conclusion regarding
the inapplicability of the $A+B$ model remains unaltered.

The precision of the SDSS MPA/JHU masses and star formation rates may
affect our conclusions. We have already mentioned the effects of
statistical uncertainties in the derived M and SFR quantities. We
repeat our work using VESPA masses and star formation rates. The
resulting values of $A$ and $B$ conform better to those found by
Sullivan et al. (2006) (see Table \ref{litcomp}), but we are still
able to reject all $A+B$ models using KS tests. See Table
\ref{vespalpha} for a summary of our results obtained using VESPA
data, and Fig. \ref{litcomp} for our best-fit values for $A$ and $B$
obtained using VESPA.

Our simplified treatment of the observation windows (\S2) could affect
the derived values of $A$ and $B$ (and $C$, where relevant) by a
common factor (which we estimate to be $<0.1$ dex). Again, this does
not affect conclusions regarding the rejection of $A+B$ models.

\subsection{Physical Implications}

The (somewhat unphysical) $SNR=A{\times}M+B{\times}SFR+C$ model is a
good fit to the data, but implies a constant rate per galaxy
independent of galaxy properties. Could the progenitors of SNe arising
from the $C$ term ($\sim68^{+27}_{-8}\%$ of the total number of SNe,
$1\sigma$ errors quoted) be due to a diffuse intergalactic stellar
population, not accounted for in MPA/JHU masses? Sand et al. (2011)
find that about 17\% of the SNe~Ia in rich clusters occur in an
intergalactic stellar population, a number that rises to almost 50\%
in groups (McGee and Balogh 2010). As discussed in Sand et al., these
numbers are in general accord with the direct intergalactic light
measurements of Gonzalez et al. (2005, 2007). If some supernovae
belong to an intergalactic population, then nearby galaxies (with
approximately the same redshifts as the supernovae) could be
incorrectly identified as hosts. A detailed modelling of this effect
is beyond the scope of this paper; however, a $C$ term that requires
$>$50\% of SNe~Ia to be intergalactic seems implausible. An alternate
interpetation of the constant $C$ model can be obtained simply by
writing the rate in the more physical form $SNR/M = A + B (SFR/M) +
C/M$. In this case $C/M$ provides some mass modulation of the $A+B$
model, which we have already noted.

We now turn to this mass dependence of the $A+B$ model, as seen
directly in Fig. \ref{fig4_2}, and also in the fits to the two
$M$-dependent models in Table \ref{ABalpha}. An effect similar to this
has been seen before in the Lick Observatory Supernova Survey (LOSS --
Li et al. 2011): SNe~Ia from LOSS possess a specific rate $SNR/M
\propto M^{-0.5}$ (rather than the expected $SNR/M \propto$
constant). In the case of LOSS, however, reliable SFRs were not
available due to a lack of spectroscopic data, rendering it difficult
to disentangle the mass-dependence of mean galaxy age or specific star
formation rate (e.g. Schiminovich et al. 2007), which could produce
some or all of this effect. In fact, a simple model incorporating ages
from Gallazzi et al. (2005) explains the LOSS data rather well
(e.g. \S 3 of Kistler et al. 2011), because low mass galaxies tend to
be younger, and have higher specific SFR's, than high mass galaxies.

Mass-dependence of the simple $A+B$ model has also been inferred by
Smith et al. (2012), who combine SDSS SN~Ia data with photometric
masses and SFR's to derive $SNR/M \propto A{\times}M^{-0.3} + B
(SFR/M)$. As already noted, however, this model fails the SFR-ranked
KS test when their parameters are assumed.

In this paper the mass-dependence of the $A+B$ model is demonstrated
in a manner that is different from, but complementary to, the
methodology of Smith et al.. First, we use
spectroscopically-determined masses and star formation rates, which
presumably should be more accurate (though resulting in a smaller
sample size). Second, we visualize the mass-dependence in a
non-parametric manner from a plot of $SNR/M$ vs. $SFR/M$ for two
different mass samples (Fig. \ref{fig4_2}). Note that this is quite
different from the $SNR/M$ vs $M$ plot in Li et al. (2011), which
includes galaxies with a wide range of SFR's, and hence may be
affected by the mass-dependence of star formation rate (though see
below). Third, we use the fact that an acceptable model must
successfully reproduce the observed distribution of supernovae with
mass and SFR, a fact that rules out most models that do not depend on
mass.

How is this mass dependence of the $A+B$ model to be interpreted? The
obvious culprit is metallicity, due to the well-established
correlation between metallicity and mass for galaxies (Tremonti et
al. 2004, Gallazzi et al. 2005 and references therein). Based on the
environmental dependence of SDSS SNR, it has been suggested by Cooper
et al.~(2009) that the SNR of galaxies may increase with decreasing
gas-phase metallicity. Kistler et al. (2011) note that for a given
initial mass, the endpoint white dwarf is more massive for lower
metallicity, a fact that may on its own (and ignoring many other
factors) lead to an enhanced SNR in low metallicity galaxies. We note
that neither of these results constitutes a direct proof of an inverse
metallicity-SNR relation. Theoretical models for {\it single
degenerate} SNe~Ia in fact suggest the opposite rate-metallicity
relation (e.g. Kobayashi et al. 2000, Langer et al. 2000).


An alternate, and somewhat more straightforward, interpretation for
the mass dependence relates to the fact
that the rate of SN~Ia from a burst of star formation (the so-called
delay time distribution or DTD) declines with time as roughly $1/t$
(Totani et al. 2008, Maoz et al. 2011). Referring to
Fig. \ref{fig4_2}, the median log mass for our low (high) mass sample
is $1.7\times10^{10}M\odot$ ($1.1\times10^{11}M\odot$). Thus, from the
mean age-mass relation of Gallazzi et al (2005), and DTD $\propto
1/t$, we would expect the rate to be about $\sim1.5-2\times$ higher in
the lower mass sample (assuming that all stars within a galaxy form at
the same epoch) -- marginally consistent with what is
observed. However, we note (Pritchet et al. 2008) that a wide range of
evolutionary models and composite stellar populations (and hence mean
stellar ages) converge to a narrow range of loci in the SNR/M
vs. SFR/M diagram in Fig. \ref{fig4_2}.

While the initial mass function (IMF) is generally assumed to be
universal, recent evidence -- both dynamical (e.g. Cappellari et
al.~2012) and spectrophotometric (e.g. Spiniello et al.~2012, Ferreras
et al.~2012) -- suggests that massive galaxies possess steeper IMF's,
with a deficit of high mass stars relative to low mass stars (see also
Kroupa et al.~2011). At a deeper level, this may well be a metallicity
effect. Since SNe~Ia are exploding WD's, and since such WD's originate
from main sequence stars as massive as 8 M$_\odot$, it follows that
low mass galaxies are more efficient (per unit galaxy mass) in
producing SNe~Ia. A detailed calculation of this effect is not yet
possible because, among other things, it requires a knowledge of the
efficiency of SN~Ia production as a function of stellar mass. However,
it is interesting to note that, for fixed total mass and a power-law
mass function $dN \propto M^\alpha dM$, varying the power $\alpha$ by
$\pm 0.5$ (as observed) around the Salpeter value of $\alpha=-2.35$
produces a change in the number of SN~Ia progenitors by a factor of
$2-4\times$ (depending on the mass range used). An IMF effect
therefore has some potential as a mechanism for the galaxy mass
dependence of SN~Ia rates.

Finally, we note the recently discovered host mass effect on SN Ia
peak luminosities (Kelly et al. 2010, Sullivan et al. 2010). A
somewhat simplistic hypothesis is that both this effect, and the host
mass modulation of supernova specific rates that we have found, are
connected. Both these effects could be caused by the presence of two
distinct classes of SN Ia progenitors, whose frequency depends on host
galaxy mass through metallicity and/or progenitor mass.

\section{Conclusions}

\begin{figure}[p]
\centering
\includegraphics[scale=0.35, angle=270]{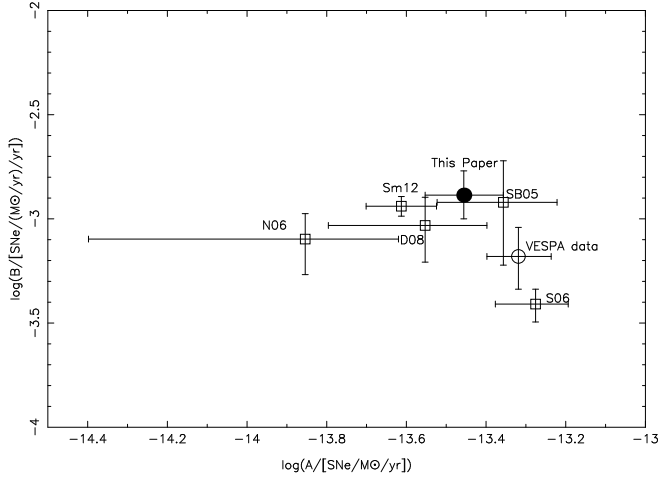}
\centering
\caption[Literature Comparison]{Results of our fits to the $A+B$ model with error bars (filled circle) plotted against other values (squares) in the literature (Scannapieco \& Bildsten 2005, Neill et al.~2006, Dilday et al.~2008, Sullivan et al.~2006, Smith et al.~2012). The open circle denotes our results obtained using VESPA data, as mentioned in the discussion. $A$ is in units of M$_\odot^{-1}$ y$^{-1}$; $B$ is in units of M$_\odot^{-1}$. \label{litcomp}}
\end{figure}

\begin{figure}[p]
\centering
\includegraphics[trim = 25mm 0mm 0mm 0mm, clip, scale=0.35]{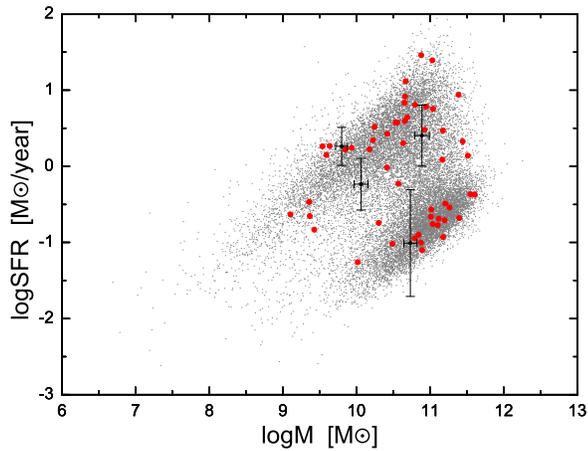}
\caption[]{Distribution of spectroscopic galaxies in the MPA/JHU catalog (grey dots), with the 53 hosted SNe (filled red circles) on a log$M$ - log$SFR$  plane. Also shown for comparison are the error bars on the masses and star formation rates of 5 randomly picked galaxies.\label{fig2_1}}
\end{figure}

\begin{figure}[p]
\centering
\includegraphics[scale=0.5, angle=270]{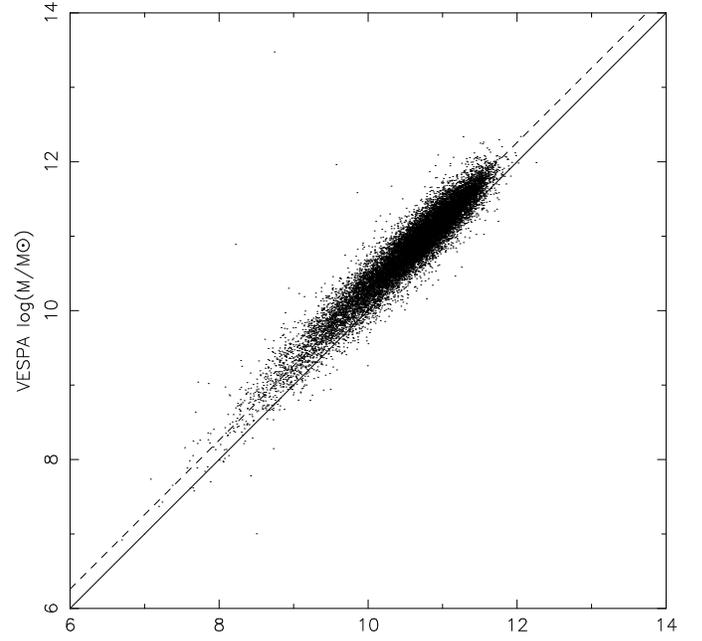}
\includegraphics[scale=0.5, angle=270]{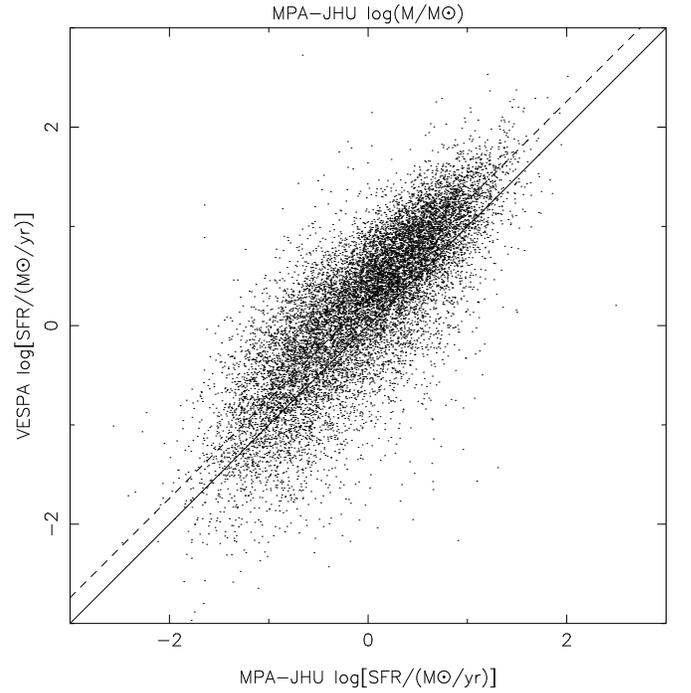}
\caption[Comparison of MPA/JHU and VESPA Masses and SFRs]{A comparison between MPA/JHU and VESPA entries for mass (top) and SFR (bottom). Note the constant offset (dashed line), as predicted in Tojeiro et al.~2009, which is the result of a calibration offset between the VESPA and MPA databases. The scatters (standard deviation) for the masses and SFRs are 0.18 dex and 0.48 dex respectively.\label{fig2_2}}
\end{figure}

\begin{figure}[p]
\centering
\includegraphics[trim = 0mm 0mm 0mm 0mm, clip, scale=0.2, angle=0]{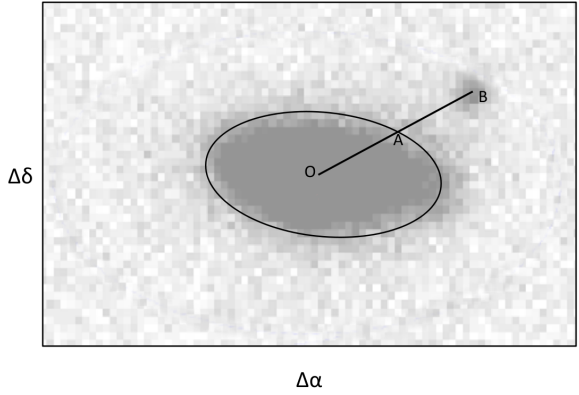}
\caption[]{Diagram illustrating the concept of $R_{25}$. The ellipse
is the 25 mag/arcsec$^{2}$ isophote for the potential host galaxy, O
is the centre, the SN~Ia is at $B$, and $A$ is the point of
intersection between the ellipse and line
OB. $R_{25}$=OB/OA. \label{R25}}
\end{figure}

\begin{figure}[p]
\centering
\includegraphics[scale=0.3, angle=270]{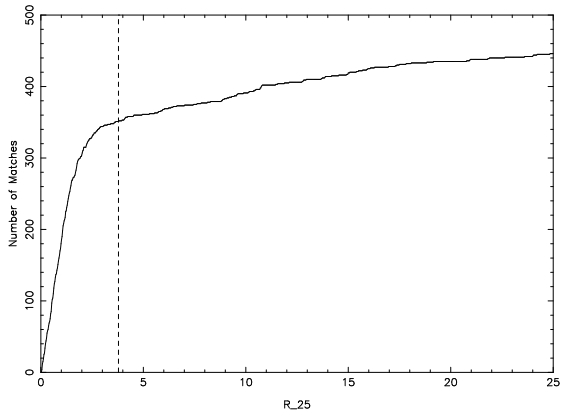}
\caption[SN Matches as a Function of Criteria]{$R_{25}$ criteria used
(x-axis) plotted against number of SNe matched (y-axis). $R_{25}=3.8$,
the criterion we use, is shown by the vertical line. Note that the
curve flattens out very rapidly beyond this line.\label{fig2_6}}
\end{figure}

\begin{figure}[p]
\centering
\includegraphics[scale=0.35, angle=270]{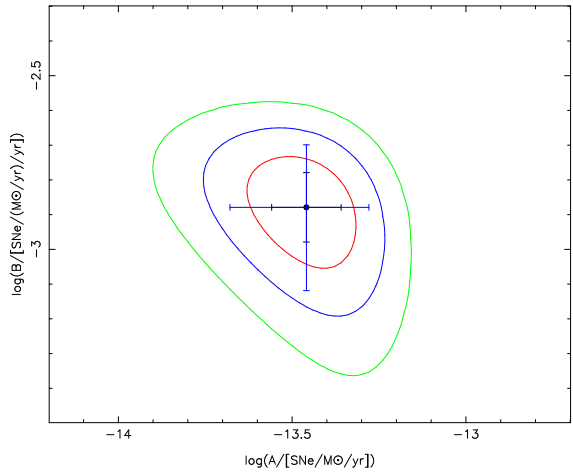}
\caption[$A+B$ Probability Contours]{Probability contours for the
values of $A$ and $B$ on an $A$-$B$ plane. The red, blue and green
contours represent 1$\sigma$, 2$\sigma$, 3$\sigma$ probability
respectively, while the circle denotes our final maximum likelihood
value. The error bars show the 1$\sigma$ and 2$\sigma$ limits of the
marginalised distributions of $A$ and $B$.\label{fig3_1}}
\end{figure}

\begin{figure}[p]
\centering
\includegraphics[scale=0.35, angle=270]{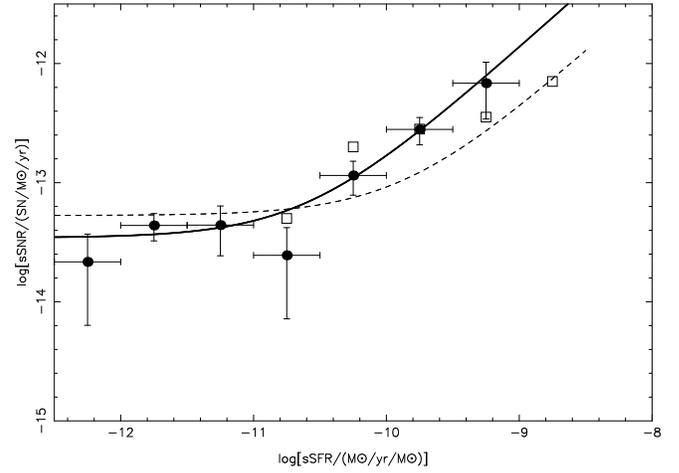}
\caption[Best-Fit $A+B$ Model vs Observed Values]{Specific supernova
rate as a function of specific star formation rate. The black points
are for our observed sample. Plotted for comparison are the
predictions of our best-fit $A+B$ model (black line), the predictions of
Sullivan et al.~2006 (dashed line), and their data (squares).\label{fig3_2}}
\end{figure}

\begin{figure}[p]
\centering
\includegraphics[trim = 15mm 0mm 0mm 0mm, clip, scale=0.35]{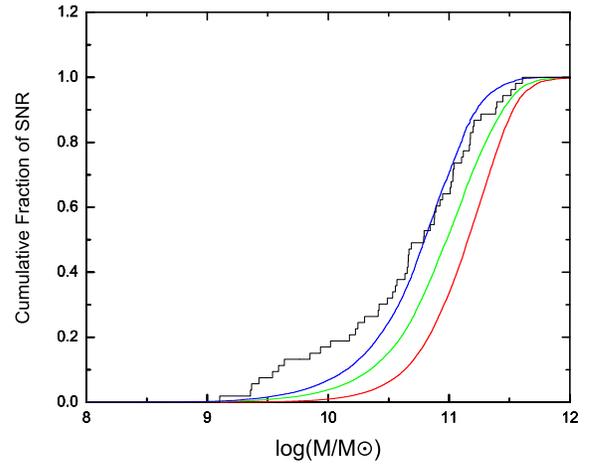}
\caption[Mass-Ranked KS tests (1)]{Cumulative distribution functions
of mass-ranked K-S tests for our best-fit $A+B$ model (green), the
$A\times$M-term-only model (red), and the $B\times$SFR-term-only model (blue). The
green and red lines are rejected by the K-S test, showing that our
best-fit $A+B$ model and all $A+B$ models with a higher $A$/$B$ ratio
do not agree with our data.\label{fig4_1}}
\end{figure}

\begin{figure}[p]
\centering
\includegraphics[trim = 15mm 0mm 0mm 0mm, clip, scale=0.35]{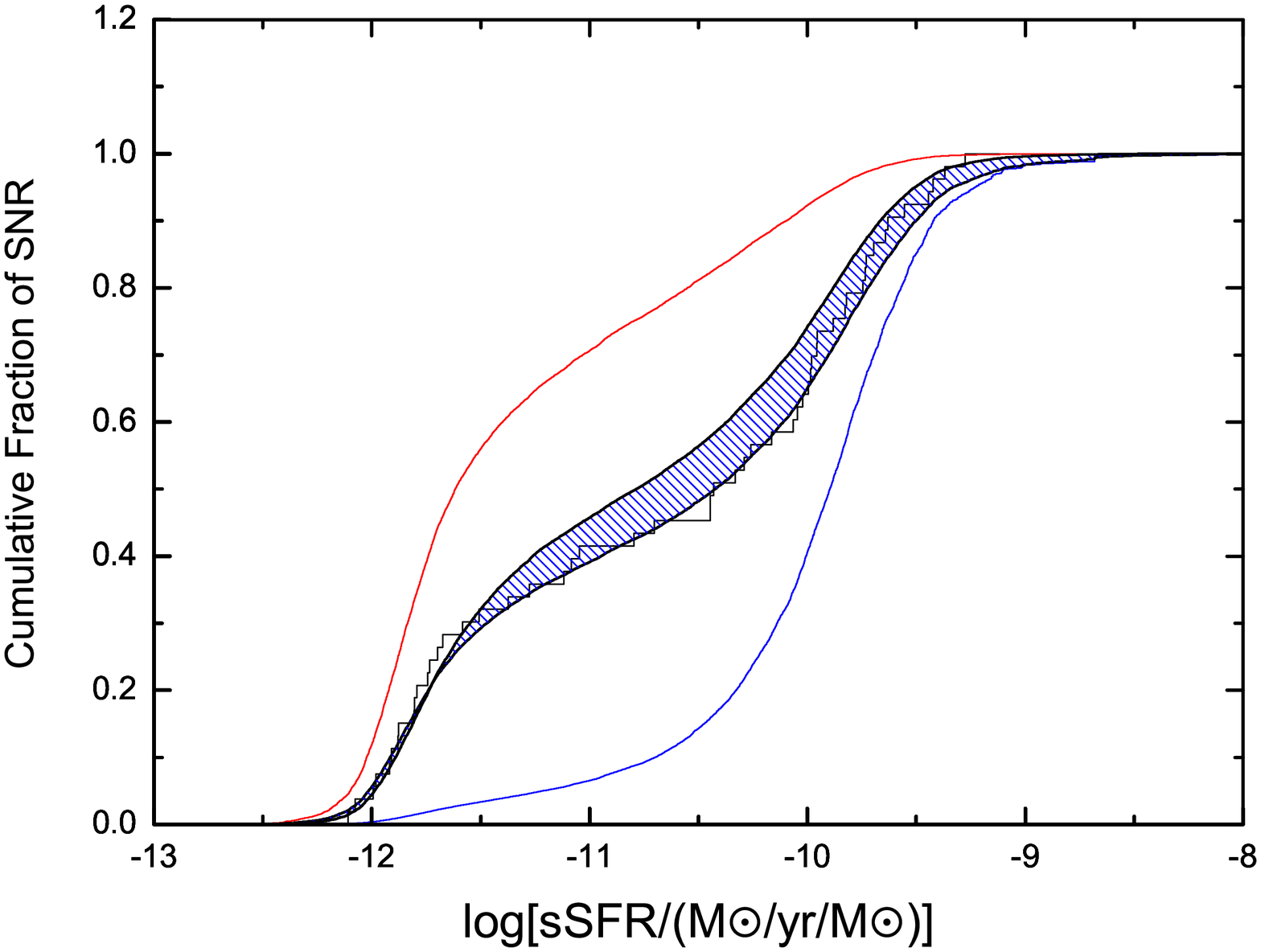}
\caption[sSFR-Ranked KS tests]{Cumulative distribution functions of
the specific-star-formation-ranked K-S tests for the various
models. The only models which were rejected were the mass-only model
(red), and the SFR-only model (blue). The remaining models
($A{\times}M+B{\times}SFR$,$A{\times}M+B{\times}SFR+C{\times}M^{-1}$,$A{\times}M+B{\times}SFR+C$,
and $(A{\times}M+B{\times}SFR)(1+C{\times}M^{-1})$) all fall within
the hashed blue area. As described in the text, this set of K-S tests
is largely inconclusive.\label{fig4_7}}
\end{figure}

\begin{figure}[p]
\centering
\includegraphics[trim = 15mm 0mm 0mm 0mm, clip, scale=0.35]{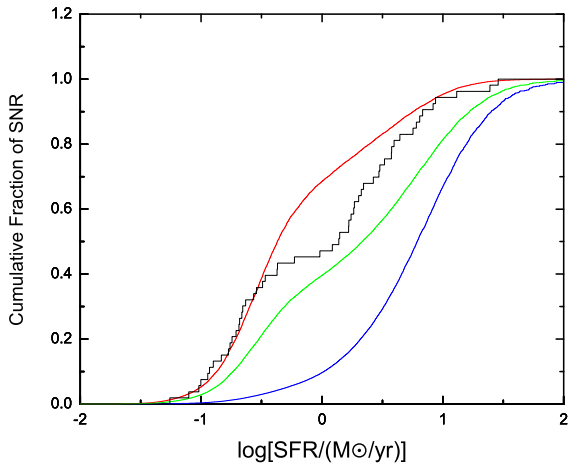}
\caption[SFR-Ranked KS tests (1)]{Cumulative distribution functions of
SFR-ranked K-S tests for our best-fit $A+B$ model (green), the
AM-term-only model (red), and the $B$SFR-term-only model (blue).  The
green and blue lines are rejected by the K-S test, showing that our
best-fit $A+B$ model and all $A+B$ models with a lower $A$/$B$ ratio
do not agree with our data. This, in conjunction with
Fig. \ref{fig4_1}, rules out all $A+B$ models.\label{fig4_8}}
\end{figure}

\begin{figure}[p]
\centering
\includegraphics[trim = 0mm 0mm 0mm 0mm, clip, scale=0.3, angle=270]{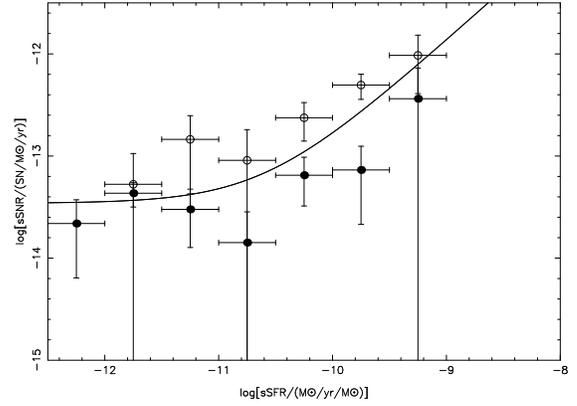}
\caption[High Mass vs Low Mass SNR Distributions on sSNR-sSFR
plane]{Specific supernova rate as a function of specific star
formation rate. The open points correspond to our low-mass sample,
and the filled points our high-mass sample. The vertical error bars
indicate Poisson uncertainties for our SNe numbers in each bin, and the
horizontal error bars correspond to bin size. The trend that the open
points tend to lie above the filled ones exists for all mass
discrimination criteria we try. Also plotted are the predictions of
our best-fit $A+B$ model. Our results have been scaled to account for
observing window issues. \label{fig4_2}}
\end{figure}

\begin{figure}[p]
\centering
\includegraphics[trim = 0mm 0mm 0mm 0mm, clip, scale=0.3, angle=270]{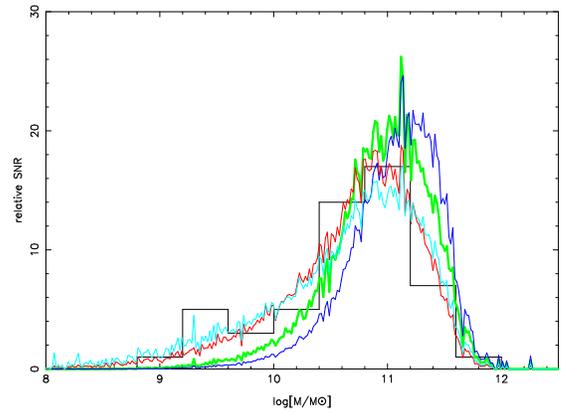}
\caption[Differential SNR Distributions (1)]{Differential distribution
functions of supernova rate as a function of host mass. The black line
corresponds to our observed data, presented as number of SNe observed
in each respective mass bin. The green and blue lines are the
predictions of our best-fit $A+B$ model and the results of Sullivan et
al.~(2006) respectively. The red line is the prediction of our
best-fit $A{\times}M+B{\times}SFR+C$ ``constant background" model, and
the cyan line corresponds to the
$(A{\times}M+B{\times}SFR)(1+C{\times}M^{-1})$ model. The predictions
have been scaled by the number of bins used within the plotted domain,
so that the lines are comparable with those of the observed data. Our
results have been scaled to account for observing window
issues.\label{fig4_3}}
\end{figure}

\begin{figure}[p]
\centering
\includegraphics[trim = 0mm 0mm 0mm 0mm, clip, scale=0.3, angle=270]{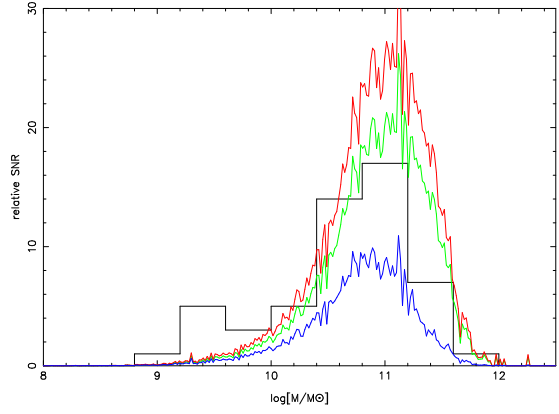}
\caption[Differential SNR Distributions (1)]{Differential distribution
functions of supernova rate as a function of host mass. The black line
corresponds to our observed data, presented as number of SNe observed
in each respective mass bin. The red and blue lines correspond to the
upper and lower 1$\sigma$ limits due to the uncertainties of our results for
the values of $A$ and $B$ respectively, plotted by setting both $A$
and $B$ to their upper and lower limits respectively. The green line
corresponds to our best-fit values. The predictions have been scaled
by the number of bins used within the plotted domain, so that the
lines are comparable with those of the observed data. Our results have
been scaled to account for observing window issues.\label{ABerr}}
\end{figure}

In summary, we have studied how Type Ia supernova rates correlate with
host galaxy masses and star formation rates. Our supernovae and host
galaxies were taken from the SDSS spectroscopic galaxy sample, which
has spectroscopic measurements of masses and star formation rates. The
reliability of the host sample was checked by comparing with VESPA
data. Our primary results are as follows:

\begin{itemize}

\item We find that maximum likelihood values
$A=3.5^{+0.9}_{-0.7}\times10^{-14}(\text{SNe/yr})(M_{\odot})^{-1}$ and
$B=1.3^{+0.4}_{-0.3}\times10^{-3}(\text{SNe/yr})(M_{\odot}$yr$^{-1})^{-1}$
provide an optimal fit to our data for a generic $A+B$ model. This
result is largely consistent with the literature (Fig. \ref{litcomp}).

\item The $A+B$ model fails to predict the mass-dependence of the
specific SNR - specific SFR relation in our sample. We also show that
\emph{no set of $A$ and $B$ can account for this discrepancy between
the model and the observed data}. This result holds regardless of any
uncertainties in the observing windows and completeness.

\item Modifications to improve the $A+B$ model are
tested. Mass-dependent models perform well in the K-S tests.

\end{itemize}

A number of explanations of these results exist. Most promising is the
galaxy mass dependence of mean stellar ages, possibly coupled with
variation in the slope of the IMF with galaxy mass. Further work is
needed to quantify the effect of these two mechanisms on supernova
rates.

\acknowledgments

This work was funded by Natural Sciences and Engineering Research
Council of Canada.

Funding for the SDSS and SDSS-II has been provided by the Alfred
P. Sloan Foundation, the Participating Institutions, the National
Science Foundation, the U.S. Department of Energy, the National
Aeronautics and Space Administration, the Japanese Monbukagakusho, the
Max Planck Society, and the Higher Education Funding Council for
England. The SDSS Web Site is http://www.sdss.org/.

The SDSS is managed by the Astrophysical Research Consortium for the
Participating Institutions. The Participating Institutions are the
American Museum of Natural History, Astrophysical Institute Potsdam,
University of Basel, University of Cambridge, Case Western Reserve
University, University of Chicago, Drexel University, Fermilab, the
Institute for Advanced Study, the Japan Participation Group, Johns
Hopkins University, the Joint Institute for Nuclear Astrophysics, the
Kavli Institute for Particle Astrophysics and Cosmology, the Korean
Scientist Group, the Chinese Academy of Sciences (LAMOST), Los Alamos
National Laboratory, the Max-Planck-Institute for Astronomy (MPIA),
the Max-Planck-Institute for Astrophysics (MPA), New Mexico State
University, Ohio State University, University of Pittsburgh,
University of Portsmouth, Princeton University, the United States
Naval Observatory, and the University of Washington.

Special thanks to Sara Ellison, Trevor Mendel and Sebastien Fabbro for
valuable discussion during the course of this work, and also for
helping with the navigation of the various databases used. We thank the referee, Alex Conley, for valuable advice concerning the data processing.

Thanks also to colleagues at Nanjing University for hardware support
during the final stages of this work.






\clearpage

\begin{center}
\begin{longtable}{ccc}

\caption{Sample of 53 Hosts for $A+B$ model fits\label{53hosts}}\\
\hline\\
   \multicolumn{1}{c}{\textbf{Host ID}} &
   \multicolumn{1}{c}{\textbf{$\log$M$^{*}$}} &
   \multicolumn{1}{c}{\textbf{$\log$SFR$^{*}$}} \\
\\
\hline
\\
\endfirsthead

\multicolumn{3}{c}{{\tablename} \thetable{} -- Continued} \\
\hline\\
   \multicolumn{1}{c}{\textbf{Host ID}} &
   \multicolumn{1}{c}{\textbf{$\log$M$^{*}$}} &
   \multicolumn{1}{c}{\textbf{$\log$SFR$^{*}$}} \\
\\
\hline
\\
\endhead

\\
\hline
  \multicolumn{3}{l}{{Continued on Next Page\ldots}} \\
\endfoot

  \\
 \hline
  \multicolumn{3}{l}{$^{*}$ M is in M$_{\odot}$, SFR is in M$_{\odot}$/yr}
\endlastfoot

  587730847691047799&   10.654&   0.834\\
  587730848501203452&    9.427&  -0.833\\
  587731173306008275&   11.118&  -0.688\\
  587731185114350067&   11.011&  -0.663\\
  587731185668849933&   11.034&  -0.760\\
  587731185669505187&   10.557&   0.575\\
  587731186195366060&   10.019&  -1.259\\
  587731186197332148&    9.936&   0.241\\
  587731511537959123&    9.590&   0.147\\
  587731511546806433&    9.850&   0.220\\
  587731513142214757&   10.489&  -1.020\\
  587731513142542420&   10.414&  -0.017\\
  587731513143328955&   10.845&  -0.898\\
  587731513146671215&   11.395&  -0.681\\
  587731513679675512&   10.226&   0.345\\
  587731513679610009&   10.534&   0.576\\
  587731513693569205&   11.017&  -0.566\\
  587731514222116993&   11.444&   0.326\\
  587731514227818648&   10.177&   0.222\\
  587731514231816228&   11.104&  -0.774\\
  587731514232209584&   11.175&   0.471\\
  587731172231872692&   11.178&  -0.930\\
  587731172233183337&   10.795&   0.809\\
  587731172767368214&   10.664&   0.919\\
  587731174914786124&   10.569&  -0.229\\
  587731185121951943&   10.669&   1.115\\
  587731185126539408&    9.637&   0.269\\
  587731185129554046&   11.168&   0.085\\
  587731185132568670&   10.878&  -1.001\\
  587731187278872773&   11.386&   0.940\\
  587731512071028897&   10.892&  -1.102\\
  587731512621465722&   10.421&   0.426\\
  587731513427624076&    9.357&  -0.470\\
  587734305949483196&   11.551&  -0.368\\
  588015507661783172&    9.366&  -0.655\\
  588015507672137965&   10.947&   0.782\\
  588015507677642829&   11.042&   0.752\\
  588015508206518484&   11.207&  -0.491\\
  588015508211368141&   10.247&   0.518\\
  588015508215431379&   11.033&   1.390\\
  588015510339256459&   11.264&  -0.541\\
  588015510339649629&   11.199&  -0.711\\
  588015510363373783&   10.661&   0.596\\
  588015508735393931&   11.513&   0.140\\
  588015509274427469&   11.608&  -0.372\\
  588015509275869191&    9.540&   0.263\\
  588015509283078244&   10.635&   0.303\\
  588015509285634176&   10.880&   1.458\\
  588015509292319354&    9.103&  -0.632\\
  588015509293760750&   10.689&   0.644\\
  588015509801599099&   10.790&  -0.946\\
  588015509811626061&   10.926&   0.479\\
  588015509814313038&   10.302&  -0.744\\
\end{longtable}

\end{center}


\begin{sidewaystable}
\begin{center}
\begin{threeparttable}
\caption{Best-Fit Parameters from Fits and Rejection Rates of Different Models\label{ABalpha}}
\begin{tabular}{ccccccc}
\hline
\multirow{2}{*}{Model} &$A$&$B$&$C$&($1-\alpha_{M}$)&($1-\alpha_{sSFR}$)&($1-\alpha_{SFR}$)\\
&$10^{-14}(M\odot)^{-1}$&$10^{-3}(M_{\odot}$yr$^{-1})^{-1}$&(appropriate units)&&&\\

\hline\\
$A{\times}M+B{\times}SFR$&$3.5^{+0.9}_{-0.7}$&$1.3^{+0.4}_{-0.3}$&-&99\%&97\%&$<$1\%\\
\\
$A{\times}M+B{\times}SFR+C{\times}M^{-1}$&$3.4^{+1.0}_{-0.6}$&$1.0^{+0.4}_{-0.2}$&$2.8^{+2.2}_{-1.5}\times10^{6}$&93\%&72\%&7\%\\
\\
$A{\times}M+B{\times}SFR+C$&$0.96^{+1.28}_{-0.95}$&$0.46^{+0.34}_{-0.26}$&$3.6^{+1.4}_{-0.4}\times10^{-3}$&2\%&34\%&22\%\\
\\
$A{\times}M$ only&$6.8{\pm}0.9$&-&-&$>$99\%&$>$99\%&$>$99\%\\
\\
$(A{\times}M+B{\times}SFR)(1+C{\times}M^{-1})$&$1.8^{+0.7}_{-1.3}$&$0.066^{+0.046}_{-0.066}$&$3.2^{+22.0}_{-2.8}\times10^{11}$&2\%&80\%&26\%\\
\\
$B{\times}SFR$ only&-&$2.7{\pm}0.4$&-&53\%&$>$99\%&$>$99\%\\
\\
\hline
\end{tabular}
\begin{tablenotes}
\footnotesize
\item ``$1-{\alpha}_{X}$" is the degree of rejection from a K-S test
in which the galaxies are ranked by ``$X$". Passing all three tests is
a necessary but insufficient condition for a model to be considered
plausible.
\end{tablenotes}
\end{threeparttable}
\end{center}
\end{sidewaystable}


\begin{table}
\begin{center}
\caption{Rejection Rates of Different Models by VESPA Data\label{vespalpha}}
\begin{tabular}{ccccc}
\hline
Model&($1-\alpha_{M}$)&($1-\alpha_{SFR}$)\\
\hline
$A{\times}M+B{\times}SFR$&$>$99\%&77\%\\
$A{\times}M+B{\times}SFR+C{\times}M^{-1}$&$>$99\%&62\%\\
$A{\times}M+B{\times}SFR+C$&39\%&42\%\\
$A{\times}M$ only&$>$99\%&$>$99\%\\
$(A{\times}M+B{\times}SFR)(1+C{\times}M^{-1})$&34\%&77\%\\
$B{\times}SFR$ only&34\%&$>$99\%\\
\hline
\end{tabular}
\end{center}
\end{table}

\end{document}